\documentclass[prc,preprint,showpacs,showkeys,nofootinbib]{revtex4}%
\usepackage{amssymb}
\usepackage{graphicx}
\usepackage{bm}
\usepackage{amsmath}
\usepackage{amsfonts}%
\newcommand{\beq}{\begin{equation}}
\newcommand{\eeq}{\end{equation}}
\newcommand{\bea}{\begin{eqnarray}}
\newcommand{\eea}{\end{eqnarray}}
\newcommand{\ben}{\begin{eqnarray*}}
\newcommand{\een}{\end{eqnarray*}}

\begin{document}
\title{Short-range correlations in two-nucleon knockout reactions}
\author{C.\,A.~Bertulani}
\email{bertulani@physics.arizona.edu}
\affiliation{Department of Physics, University of Arizona, Tucson, Arizona 85721}
\date{\today}

\begin{abstract}
A theory of short-range correlations in two-nucleon removal due to elastic
breakup (diffraction dissociation) on a light target is developed.
Fingerprints of these correlations will appear in momentum distributions of
back-to-back emission of the nucleon pair. Expressions for the momentum
distributions are derived and calculations for reactions involving stable and
unstable nuclear species are performed. The signature of short-range
correlations in other reaction processes is also studied.

\end{abstract}
\pacs{21.10.Jx,24.50.+g, 25.60.-t, 27.20.+n}
\keywords{Knockout reactions, short-range correlations, momentum distributions}
\maketitle

\section{Introduction}

A primary goal of nucleus-nucleus scattering has been to learn about nuclear
structure. This has become even more critical in recent years, when many
groups became very active in the investigation of the physics of nuclei far
from the stability, mainly using nucleus-nucleus scattering processes at
intermediate energies ($E_{lab}\simeq100$ MeV/nucleon). The theoretical
complexity of such collisions has given rise to the use of a number of
different approximations. The adequate theoretical tool for this purpose is
Glauber's multiple-scattering theory \cite{Gl59}. It has long been known both
for its simplicity and amazing predictive power. One can find copious examples
in the literature where the Glauber theory allows for a simple physical
interpretation of experimental results as well as their quantitative analysis
\cite{FH80,HN81,BHM02}. In fact, fragmentation reactions of the type discussed
here have already been successfully analyzed in the framework of Glauber's
theory: in one-nucleon-removal reactions, the momentum distribution of the
outgoing fragment has been shown to reflect the momentum distribution of the
nucleon which is removed from the surface of the projectile nucleus
\cite{HN81}. \ However, because of complications involving multiple scattering
processes in nucleus-nucleus collisions, a full Glauber multiple scattering
expansion is impracticable. Fortunately, the study of many direct nuclear
processes, e.g. nucleon knockout, or stripping, elastic breakup (diffraction
dissociation), etc, are possible using the optical limit of the Glauber
theory, in which the nuclear ground-state densities and the nucleon-nucleon
total cross sections are the main input. In fact, this method has become one
of the main tools in the study of nuclei far from stability \cite{HT04}. When
departures from the optical limit are observed, multiple\ nucleon-nucleon
collisions and in-medium effects of the nucleon-nucleon interaction and
nucleon-nucleon correlations become relevant.

Very peripheral collisions, with impact parameters just around the sum of the
nuclear radii (grazing collisions), or larger, are well established tools for
studying nuclear properties with intermediate energies and relativistic heavy
ion collisions \cite{BB88,Gl98,BP99}. These collisions lead to excitation of
giant resonances through both electromagnetic and strong interactions. At
intermediate energy collisions ($E_{lab}\simeq100$ MeV/nucleon), or higher,
the collision\ time is short and the action of the short-range nuclear
interaction can excite the surface region of the colliding nuclei. This
excitation can equilibrate forming a compound nucleus, and/or give rise to
pre-equilibrium emission or other fast dissipation processes.

An interesting reaction mechanism in high-energy peripheral nucleus-nucleus
collisions was suggested by Feshbach and Zabek \cite{FZ77,Fes80}. This
mechanism has been applied in refs. \cite{BD78,Kin86,TD81,TDM84,NDT85,DNT86}
to the calculation of pion production in heavy ion collisions from
subthreshold to relativistic energies. It is assumed that pions are produced
in peripheral processes through the excitation of the projectiles to a
$\Delta$-isobar giant resonance. The results of these calculations were
compared to inclusive pion production data for incident energies from 50 MeV
to 2 GeV per nucleon. As emphasized by those authors, this comparison is not
very meaningful at high energy where peripheral processes are expected to
contribute very little to the total pion production. However, at subthreshold
energies, coherent pion production should dominate the cross section. This
mechanism is known as the nuclear Weizsaecker-Williams method. It works as follows.

The uncertainty relation associated to the variation of the time-dependent
nuclear field on a scale $\Delta z$ leads a relation between the  energy,
$\Delta E$, and momentum transfer, $\Delta p$:%
\[
\Delta E\simeq\frac{\hbar}{\Delta t}=\frac{\hbar\mathrm{v}}{\Delta
z},\ \ \ \ \ \Delta p\simeq\frac{\hbar}{\Delta z}\ \ \ \ \Longrightarrow
\ \ \Delta E=\mathrm{v}\Delta p.
\]
The last equation on the right is the dispersion relation of a phonon. For
typical situations, $\Delta z$ is a few fermis and the nuclear interaction
pulse carries several hundred MeV. This relation can also be directly obtained
from the collision kinematics. Let $\left(  E_{i},\mathbf{p}_{i}\right)  $ be
the initial momentum of the projectile and $\left(  \Delta E,\Delta
\mathbf{p}\right)  $ the energy-momentum transfer in the reaction. One has%
\[
\mathbf{P}_{f}=\mathbf{P}_{i}-\Delta\mathbf{p,}\ \ \ \ \ \ \ \ \ E_{f}%
=E_{i}-\Delta E.
\]
From these relations one finds%
\[
\frac{\mathbf{P}_{i}.\Delta\mathbf{p}}{E_{i}}-\Delta E=\frac{-\left(  \Delta
E\right)  ^{2}+\left(  \Delta p\right)  ^{2}+\left(  M_{i}^{2}-M_{f}%
^{2}\right)  c^{4}}{2E_{i}}.
\]
Neglect the term on the right-hand side, one gets%
\begin{equation}
\Delta E=\mathbf{v\cdot\Delta p}=\mathrm{v}\Delta p_{z},\label{phonon}%
\end{equation}
where $\Delta p_{z}$\ is the momentum transfer along the longitudinal direction.

The above relation can only be satisfied for nuclear excitations of very small
momentum transfers, even for moderately large energy transfers. This is the
case for the excitation of giant resonances. Thus, the nuclear interaction in
grazing nuclear collisions is an effective tool to probe giant resonances (for
a review see, e.g. ref. \cite{CF95}). For very large impact parameters (larger
than the sum of the nuclear density radii) only the electromagnetic
interaction is present, and eq. \ref{phonon} (with $v\simeq c$) is just the
energy-momentum relation of a real photon. In fact, relativistic Coulomb
excitation is another useful tool for investigating giant resonances
\cite{BB88,BP99}.

The phonon-like relation, eq. \ref{phonon}, is also a tool for studying
nucleon-nucleon short-range correlations. The energy in eq. \ref{phonon} could
hardly be absorbed by a single nucleon since it would carry the momentum
$\sim\sqrt{2m\Delta E}$, which is appreciably larger than that of eq.
\ref{phonon}. However, the phonon could be absorbed by a correlated
nucleon-pair, which can have large kinetic energy and small total momentum,
when the nucleons move in approximately opposite directions. This mechanism
has been exploited by previous authors to study the emission of correlated
pairs in relativistic heavy ion collisions \cite{BCD89,NN91}. Remarkably,
refs. \cite{FZ77,Fes80} do not treat properly the nuclear absorption at small
impact parameters, leading to very large cross sections for the emission of
correlated pairs in peripheral collisions.

In many-body physics the word correlation is used to indicate effects beyond
mean-field theories. In nuclear physics one distinguishes between short- and
long-range correlations. Nuclear collective phenomena such as vibrations and
rotations are known to be ruled by long-range correlations. These effects are
relatively well known. Short-range correlations is also a subject of intensive
studies in nuclear physics (see, e.g.
\cite{Fra81,Fra88,Dim00,BD02,Tan03,Ryc04}). The sources of short-range
correlations are the strong repulsive core of the microscopic nucleon-nucleon
interaction at short internucleon distances. The nucleon-nucleon interaction
becomes strongly repulsive at short distances. The phase shifts for $^{1}%
$S$_{0}$ and $^{3}$S$_{1}$ are positive at low, and become negative at higher
energies \cite{MAW69}. This indicates a repulsive core at short distances and
attraction at long distances. In the nuclear medium this repulsive interaction
is strongly influenced by Pauli blocking. The search for nuclear phenomena
showing short-range correlations effects is one of the most discussed topics
in the nuclear structure community. For the nuclear reaction community, the
importance of Pauli correlations in high energy nucleus-nucleus collisions has
prompted the consideration of effects of dynamical short-range correlations.
When one treats nucleus-nucleus collisions at high energies with an optical
phase shift function one can include both the center-of-mass correlations and
two-body correlations in a straightforward manner to obtain a rapidly
converging series for the physical observables.

It would be proper at this time to look for fingerprints of
short-range correlations in high-energy collisions involving rare
nuclear isotopes. Recent experiments on knockout reactions seem to
indicate a quenching of the spectroscopic factor relative to
shell-model predictions in neutron-rich nuclei \cite{HT04}. \ This
reduction is thought to be a consequence of short-range correlations
which spread the single particle strength to states with higher
energies. In fact, systematic studies with the $A\left(
e,e^{\prime}p\right)  $ reaction have provided ample evidence for
this quenching phenomenon \cite{Pan97}. In this context, two-proton
knockout reactions with exotic nuclear beams seem to be a promising
tool to investigate short-range correlations in neutron(proton)-rich
nuclei \cite{Baz02}. Indeed, for decades two-proton knockout has
been considered a valuable tool to study short-range correlations in
proton-nucleus and electron-nucleus processes (for recent work, see
e.g. \cite{Tan03,Ryc04}). In high-energy nucleus-nucleus collisions,
the phonon mechanism, proposed by Feshbach and Zabek, is a useful
guide for the investigation of short-range correlations.

The plan of this paper is as follows. In this work we treat the effects of
short-range correlations on heavy-ion scattering at high energies. In Sec. 2
the Glauber formalism for diffraction dissociation is reviewed. In section 3
this formalism is shown to lead to the same result as the traditional DWBA
calculations under the proper conditions. This is an important point, as
diffraction dissociation and DWBA approaches are commonly referred to as
distinct reaction mechanisms in the literature. In section 4 the role of
absorption and Lorentz boosts is discussed. In section 5 the formalism is
applied to heavy-ion collisions in the presence of two-body correlations,
showing the connection with the Feshbach and Zabek method. The significance of
short-range correlations is further discussed. In sec. 6 the formalism is
applied to carbon-carbon and $^{11}\mathrm{Li}+^{9}\mathrm{Be}$ collisions. In
Sec. 7 some concluding remarks are made.

\section{Diffraction dissociation}

Let us consider high energy scattering, so that the energy transfer in the
collision, $\Delta E$, is much smaller than the kinetic energy of the
colliding nuclei, $E$. In most cases, one is also interested in processes for
which the fragments fly in the forward direction, i.e. we will also assume
that $\Delta\theta\ll1$. In such situations the particle wavefunctions are
well described by eikonal waves \cite{BD04}, i.e. a plane wave distorted by an
interaction, $V$, so that the $S$-matrix is given by the simple formula
$S\left(  b\right)  =\exp\left[  -(i/\hbar\mathrm{v})\int dZ\ V\left(
R\right)  \right]  $, with $\mathrm{v}$ equal to the projectile velocity and
$R=\left(  \mathbf{b},Z\right)  $ the distance between projectile and target
($V$ is assumed to be spherically symmetric). Extending this approach to
account for scattering of bound particles, the initial and final states are
given by%
\begin{equation}
\Psi_{i}=\phi_{i}\left(  \mathbf{r}\right)  \exp\left(  i\mathbf{k}%
\cdot\mathbf{R}\right)  ,\ \ \ \ \ \ \ \ \ \ \ \ \Psi_{f}=\phi_{f}\left(
\mathbf{r}\right)  S\left(  b\right)  \ \exp\left(  i\mathbf{k}\cdot
\mathbf{R}\right)  \ ,
\end{equation}
where $\phi_{i,f}\left(  \mathbf{r}\right)  $ are the initial and final
probability amplitudes (wavefunctions) that a particle in the projectile is at
a distance $\mathbf{r}$ from the center of mass. The particle's $S$-matrix,
$S\left(  b\right)  $, accounts for the distortion due to the interaction.

For a projectile with two-body structure (e.g. a core+valence particle)%
\begin{align}
\Psi_{i} &  =\phi_{i}\left(  \mathbf{r}\right)  \exp\left[  i\left(
\mathbf{k}_{c}\cdot\mathbf{r}_{c}+\mathbf{k}_{v}\cdot\mathbf{r}_{v}\right)
\right]  \nonumber\\
\Psi_{f} &  =\phi_{f}\left(  \mathbf{r}\right)  S_{c}\left(  b_{c}\right)
S_{v}\left(  b_{v}\right)  \exp\left[  i\left(  \mathbf{k}_{c}^{\prime}%
\cdot\mathbf{r}_{c}+\mathbf{k}_{v}^{\prime}\cdot\mathbf{r}_{v}\right)
\right]  \ ,\label{eikw1}%
\end{align}
where now $\phi_{i,f}\left(  \mathbf{r}\right)  $ are the initial and final
intrinsic wavefunctions of the (core+valence particle) as a function of
\ $\mathbf{r=r}_{1}-\mathbf{r}_{2}$. The relation between the intrinsic,
$\mathbf{r}$, and center of mass, $\mathbf{R}$, coordinates is given in terms
of the mass ratios $\beta_{i}=m_{i}/m_{P}$. Explicitly, $\mathbf{r}%
_{v}=\mathbf{R}+\beta_{c}\mathbf{r}$ and $\mathbf{r}_{c}=\mathbf{R}-\beta
_{v}\mathbf{r}$. The core and valence particle $S$-matrices, $S_{c}\left(
b_{c}\right)  $ and$\ S_{v}\left(  b_{v}\right)  $, account for the distortion
due to the interaction with the target.

The probability amplitude for diffraction dissociation is the overlap between
the two wavefunctions above, i.e.
\begin{equation}
A_{\mathrm{(diff)}}=\int d^{3}r_{c}d^{3}r_{v}\ \phi_{f}^{\ast}\left(
\mathbf{r}\right)  \phi_{i}\left(  \mathbf{r}\right)  \delta\left(
z_{c}+z_{v}\right)  S_{c}\left(  b_{c}\right)  S_{v}\left(  b_{v}\right)
\exp\left[  i\left(  \mathbf{q}_{c}\cdot\mathbf{r}_{c}+\mathbf{q}_{v}%
\cdot\mathbf{r}_{v}\right)  \right]  ,\label{dif1}%
\end{equation}
where $\mathbf{q}_{c}=\mathbf{k}_{c}^{\prime}-\mathbf{k}_{c}$ is the momentum
transfer to the core particle, and accordingly for the valence particle. The
above formula yields the probability amplitude that the\ projectile starts the
collision in a bound state and ends up as two separated pieces.\ The
$S$-matrices, $S_{c}$ and $S_{v}$ carry all the information about the
dissociation mechanism. The delta-function $\delta\left(  Z\right)  $\  in eq.
\ref{dif1} was introduced to account for the fact that the $S$-matrices
calculated in the eikonal approximation only depend on the transverse direction.

It is instructive to follow another argument to obtain eq. \ref{dif1}. If only
the core scatters elastically, whereas the valence particle remains in its
unaltered plane wave state, the final projectile wavefunction is given by
\begin{equation}
\Psi_{f}^{\mathrm{(scatt)}}=\phi_{f}\left(  \mathbf{r}\right)  \ \left[
1-S_{c}\left(  b_{c}\right)  \right]  \ \exp\left[  i\left(  \mathbf{k}%
_{c}^{\prime}\cdot\mathbf{r}_{c}+\mathbf{k}_{v}^{\prime}\cdot\mathbf{r}%
_{v}\right)  \right]  .
\end{equation}

The factor $\left[  1-S_{c}\left(  b_{c}\right)  \right]  $ is the amplitude
for elastic scattering of the core. The same relation can be applied for the
valence particle. The diffraction dissociation occurs by subtracting the
simultaneous scattering of the core+valence particle, represented by $\left[
1-S_{c}\left(  b_{c}\right)  \right]  \left[  1-S_{v}\left(  b_{v}\right)
\right]  ,$ from the independent scattering of core and the valence particle,
i.e.%
\begin{equation}
\widehat{S}_{\mathrm{(diff)}}=\left[  1-S_{c}\left(  b_{c}\right)  \right]
\left[  1-S_{v}\left(  b_{v}\right)  \right]  -\left[  1-S_{c}\left(
b_{c}\right)  \right]  -\left[  1-S_{v}\left(  b_{v}\right)  \right]
=S_{c}\left(  b_{c}\right)  S_{v}\left(  b_{v}\right)  -1.
\end{equation}
The factor (-1) is not relevant because of the orthogonality of the
wavefunctions $\phi_{i}\left(  \mathbf{r}\right)  $ and $\phi_{f}\left(
\mathbf{r}\right)  $. Using $A_{\mathrm{(diff)}}=\left\langle \phi
_{i}\ \varphi_{\mathbf{k}_{1},\mathbf{k}_{2}}\left\vert \widehat
{S}_{\mathrm{(diff)}}\right\vert \phi_{f}\ \varphi_{\mathbf{k}_{1}^{\prime
},\mathbf{k}_{2}^{\prime}}\right\rangle $, with $\varphi_{\mathbf{k}%
_{1},\mathbf{k}_{2}}$ equal to plane waves, we regain eq. \ref{dif1}. We thus
see that diffractive dissociation (or elastic nuclear breakup) arises from the
momentum transfer to each particle due to elastic scattering, subtracting the
momentum transfer to their center of mass.

The cross section for the diffraction process $\phi_{i}\left(  \mathbf{r}%
\right)  \rightarrow\phi_{f}\left(  \mathbf{r}\right)  $ is given by%
\begin{equation}
d\sigma=\rho\left(  E\right)  \left\vert \int d^{3}r_{c}\ d^{3}r_{v}\ \phi
_{f}^{\ast}\left(  \mathbf{r}\right)  \phi_{i}\left(  \mathbf{r}\right)
\delta\left(  z_{c}+z_{v}\right)  S_{c}\left(  b_{c}\right)  S_{v}\left(
b_{v}\right)  \exp\left[  i\left(  \mathbf{q}_{c}\cdot\mathbf{r}%
_{c}+\mathbf{q}_{v}\cdot\mathbf{r}_{v}\right)  \right]  \right\vert
^{2},\label{dif}%
\end{equation}
where $\rho\left(  E\right)  \ $is the density of final states, $\rho\left(
E\right)  =\delta\left(  Q_{z}\right)  d^{3}q_{c}d^{3}q_{v}/\left(
2\pi\right)  ^{5}$, where $\mathbf{Q}=\mathbf{q}_{c}+\mathbf{q}_{v}$ is the
momentum transfer to the center of mass of the projectile. The delta function
accounts for the conservation of the longitudinal momentum of the projectile
arising from the use of eikonal wavefunctions (i.e. no dependence on the
longitudinal c.m. scattering).

It is important to notice that the above formula is somewhat different than
eq. 8\ of ref. \cite{HEB96}\ . In that reference the coordinates $\mathbf{r}%
$,$\ \mathbf{R}$ were used from the start. One can transform the integral of
eq. \ref{dif} to those variables. The Jacobian of the transformation is equal
to one and $d^{3}r_{c}d^{3}r_{v}=d^{3}rd^{3}R$, $d^{3}q_{c}d^{3}q_{v}%
=d^{3}qd^{3}Q$, where $\mathbf{q}=\beta_{c}\mathbf{q}_{v}-\beta_{v}%
\mathbf{q}_{c}$ is the momentum transfer to the intrinsic coordinates of the
projectile.\ Thus, in the coordinates $\mathbf{r}$,$\ \mathbf{R,}$ \ eq.
\ref{dif} reduces to%
\begin{equation}
d\sigma=\frac{d^{3}qd^{2}Q}{\left(  2\pi\right)  ^{5}}\left\vert \int
d^{3}rd^{2}b\ \phi_{f}^{\ast}\left(  \mathbf{r}\right)  \phi_{i}\left(
\mathbf{r}\right)  S_{c}\left(  b_{c}\right)  S_{v}\left(  b_{v}\right)
\exp\left[  i\left(  \mathbf{q}\cdot\mathbf{r}+\mathbf{Q}\cdot\mathbf{b}%
\right)  \right]  \right\vert ^{2}.\label{dif2}%
\end{equation}

The above formula reduces to eq. 8 of ref. \cite{HEB96} if one sets $\beta
_{v}=1$ and $\beta_{c}=0$. \ In this equation, $\phi_{f}\left(  \mathbf{r}%
\right)  $ can be taken as any final state of the projectile. Thus, it is not
only appropriate to calculate \textit{diffraction dissociation}, but also
\textit{diffraction excitation}. Diffraction excitation occurs when the final
state $\phi_{f}\left(  \mathbf{r}\right)  $ is a bound state. If it is a state
in the continuum (diffraction dissociation), then $\phi_{f}\left(
\mathbf{r}\right)  $ should be set to the unity\footnote{Neglecting final
state interactions. If final state interactions are important, $\phi
_{f}\left(  \mathbf{r}\right)  $ is the distortion correction to the plane
wave. }, since the part of the wavefunction given by $S_{c}S_{v}\exp\left[
i\left(  \mathbf{k}_{c}^{\prime}\cdot\mathbf{r}_{c}+\mathbf{k}_{v}^{\prime
}\cdot\mathbf{r}_{v}\right)  \right]  $ already accounts for the proper
wavefunction of the projectile. A natural improvement of eq. \ref{dif} is to
include final state interactions between the core and the valence particle in
the coordinate dependence of $\phi_{f}\left(  \mathbf{r}\right)  .$

Since I claim here that eq. \ref{dif2} can also be used for calculating
excitation cross sections, it is adequate to show the relation of this
approach to the traditional DWBA and semiclassical methods for nuclear
excitation in nucleus-nucleus collisions. We will see that the latter are
perturbative expansions of the eq. \ref{dif2}.

\section{DWBA and semiclassical methods}

On can factorize the $S$-matrices defined in section 2 for the interaction of
the core and valence particle with the target in terms of their phase-shifts
\begin{equation}
\chi=-\frac{i}{\hbar v}\int_{-\infty}^{\infty}dZ\ V\left(  R\right)
\ .\label{eikphase}%
\end{equation}
In the weak interaction limit, or perturbation limit, the phase-shifts are
very small so that%
\begin{align}
S_{c}\left(  b_{c}\right)  S_{v}\left(  b_{v}\right)   &  =\exp\left[
i\left(  \chi_{c}+\chi_{v}\right)  \right]  \ \simeq1+i\chi_{c}+i\chi
_{v}\nonumber\\
&  =1+\frac{1}{\hbar v}\int V_{cT}\left(  \mathbf{r}_{c}\right)
\ dz_{c}+\frac{1}{\hbar v}\int V_{vT}\left(  \mathbf{r}_{v}\right)
\ dz_{v}.\label{dwba1}%
\end{align}

The factor 1 does not contribute to the breakup\textbf{. }Thus, inserting the
result above in eq. \ref{dif1}, one obtains%
\begin{equation}
A_{\mathrm{(PWBA)}}\simeq\frac{1}{\hbar v}\int d^{3}r_{c}d^{3}r_{v}\ \phi
_{f}^{\ast}\left(  \mathbf{r}\right)  \phi_{i}\left(  \mathbf{r}\right)
\left[  V_{cT}\left(  \mathbf{r}_{c}\right)  +V_{vT}\left(  \mathbf{r}%
_{v}\right)  \right]  \exp\left[  i\left(  \mathbf{q}_{c}\cdot\mathbf{r}%
_{c}+\mathbf{q}_{v}\cdot\mathbf{r}_{v}\right)  \right]  ,\label{dwba2}%
\end{equation}
where the integrals over $z_{c}$ and $z_{v}$ in eq. \ref{dwba1} were absorbed
back to the integrals over $\mathbf{r}_{c}$ and $\mathbf{r}_{v}$ after use of
the delta-function $\delta\left(  z_{c}+z_{v}\right)  $. The above equation is
nothing more than the plane-wave Born-approximation (PWBA) amplitude. However,
absorption is not treated properly. For small values of $\mathbf{r}_{c}$ and
$\mathbf{r}_{v}$ the phase-shifts are not small and the approximation used in
eq. \ref{dwba1} fails.\ A better approximation is to assume that for small
distances, where absorption is important, $S_{c}\left(  b_{c}\right)
S_{v}\left(  b_{v}\right)  \simeq S\left(  b\right)  $, where the right-hand
side is the $S$-matrix for the projectile scattering as a whole on the target.
Using the coordinates $\mathbf{r}$ and$\ \mathbf{R}$ , and defining
$U_{int}(\mathbf{r,R})=V_{cT}\left(  \mathbf{r}_{c}\right)  \ +V_{nT}\left(
\mathbf{r}_{n}\right)  $, one gets
\begin{equation}
T_{\mathrm{(DWBA)}}=\hbar vA_{\mathrm{(DWBA)}}\simeq\int d^{3}rd^{3}%
R\ \phi_{f}^{\ast}\left(  \mathbf{r}\right)  \exp\left[  i\mathbf{q}%
\cdot\mathbf{r}\right]  \phi_{i}\left(  \mathbf{r}\right)  U_{int}%
(\mathbf{r,R})S\left(  b\right)  \exp\left[  i\mathbf{Q}\cdot\mathbf{R}%
\right]  \ .\label{TDWBA0}%
\end{equation}

In elastic scattering, or excitation of collective modes (e.g. giant
resonances), the momentum transfer to the intrinsic coordinates can be
neglected and the equation above can be written as%
\begin{equation}
T_{\mathrm{(DWBA)}}=\left\langle \chi^{\left(  -\right)  }\left(
\mathbf{R}\right)  \phi_{c}\left(  \mathbf{r}\right)  \left\vert
U_{int}(\mathbf{r,R})\right\vert \chi^{\left(  +\right)  }\left(
\mathbf{R}\right)  \phi_{i}\left(  \mathbf{r}\right)  \right\rangle \ ,
\label{TDWBA}%
\end{equation}
which has the known form of the DWBA T-matrix. The scattering phase space now
only depends on the center of mass momentum transfer $\mathbf{Q}$. When the
center of mass scattering waves are represented by eikonal wavefunctions, one
has%
\begin{equation}
\chi^{\left(  -\right)  \ast}\left(  \mathbf{R}\right)  \chi^{\left(
+\right)  }\left(  \mathbf{R}\right)  \simeq S\left(  b\right)  \exp\left[
i\mathbf{Q}\cdot\mathbf{R}\right]  \ . \label{chieik}%
\end{equation}
This shows that the PWBA and the DWBA are perturbative expansions of the
diffraction dissociation formula \ref{dif1}.

In DWBA (or in the eikonal approximation, eq. \ref{chieik}), $b$ does not have
the classical meaning of an impact parameter. To obtain the semiclassical
limit one goes one step further. By using eq. \ref{TDWBA0} and assuming that
$R$ depends on time so that $R=\left(  \mathbf{b},z=vt\right)  $, the
semiclassical scattering amplitude is given by $A_{\mathrm{(semiclass)}%
}^{\left(  i\rightarrow f\right)  }=i\int d^{2}b\ a_{\mathrm{(semiclass)}%
}^{\left(  i\rightarrow f\right)  }\left(  b\right)  $ exp$\left(
i\mathbf{Q}\cdot\mathbf{b}\right)  $, where%
\begin{equation}
a_{\mathrm{(semiclass)}}^{\left(  i\rightarrow f\right)  }\left(  b\right)
=\frac{1}{i\hbar}\ S\left(  b\right)  \int dtd^{3}r\ \exp\left(  i\omega
_{if}\ t\right)  \phi_{f}^{\ast}\left(  \mathbf{r}\right)  U_{int}%
(\mathbf{r,}t)\phi_{i}\left(  \mathbf{r}\right)  \ ,\label{semi1}%
\end{equation}
where eq. \ref{phonon} was used ($Q_{z}Z=\omega_{if}\ t$).

The semiclassical probability for the transition $\left(  i\rightarrow
f\right)  $\ is obtained from the above equation after squaring it of
integrating it over \textbf{Q}. One gets $\sigma^{\left(  i\rightarrow
f\right)  }=\int d^{2}b\ P_{\mathrm{(semiclass)}}^{\left(  i\rightarrow
f\right)  }\left(  b\right)  $, where $P_{\mathrm{(semiclass)}}^{\left(
i\rightarrow f\right)  }\left(  b\right)  =\left\vert a_{\mathrm{(semiclass)}%
}^{\left(  i\rightarrow f\right)  }\left(  b\right)  \right\vert ^{2}$, with
$b$ having now the explicit meaning of an impact parameter. Thus,
$a_{\mathrm{(semiclass)}}^{\left(  i\rightarrow f\right)  }\left(  b\right)
$, is the semiclassical excitation amplitude. Equation \ref{semi1} is
well-known (for example in Coulomb excitation at low energies) except that the
factor $S\left(  b\right)  $ is usually set to one. In high energy collisions
it is crucial to keep this factor, as it accounts for refraction and
absorption at small impact parameters: \ $\left\vert S\left(  b\right)
\right\vert ^{2}=\exp\left[  2\chi^{\mathrm{(imag)}}\right]  $, where
$\chi^{\mathrm{(imag)}}$ is calculated with the imaginary part of the optical
potential. The derivation of the DWBA and semiclassical limits of eikonal
methods can be easily extended to higher-orders in the perturbation $V$. The
eikonal method includes all terms of the perturbation series in the
sudden-collision limit.

\begin{figure}[ptb]
\begin{center}
\includegraphics[
height=2.3722in,
width=3.0675in
]{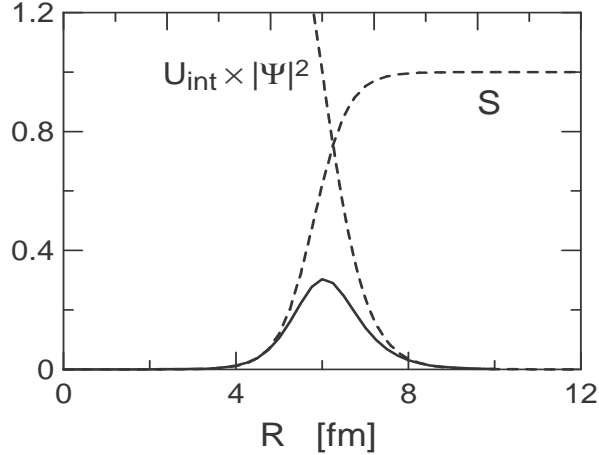}
\end{center}
\caption{Schematic diagram showing how the product of the S-matrix and the
interaction potential (weighted by the wavefunction) would limit the cross
section to grazing impact parameters.}%
\label{SXU}%
\end{figure}

\section{Role of absorption and of Lorentz boosts}

At this point it is interesting to consider the calculation performed by
Feshbach and Zabek \cite{FZ77}. In that work, eq. \ref{semi1}, or its
equivalent PWBA form, eq. \ref{dwba2}, without a proper account of the strong
absorption at small impact parameters (described in eq. \ref{semi1} by
$S\left(  b\right)  $),\ was used to calculate the total cross section for
emission of a correlated nucleon pair in peripheral collisions with heavy
ions. Also, interactions without imaginary parts were used. As a consequence,
they found extremely large cross sections; $\sim1$ barn for $^{16}$O+$^{16}$O
collisions at energies $\sim1$ GeV/nucleon. This is certainly inconsistent
with perturbation theory. As seen schematically in figure \ref{SXU}, the
product of the $S$-matrix and the interaction potential implies that the
reaction occurs in a narrow region at \textquotedblleft grazing" impact
parameters. The width of this region is approximately $\Delta\simeq1-2$ fm.
The cross section might be written as $\sigma\simeq2\pi\Delta\left(
R_{P}+R_{T}\right)  P$, where $P$ is the average probability for this reaction
to occur within the impact parameter interval $\Delta$, and $R_{P}$ $\left(
R_{T}\right)  $ is the projectile (target) radius. For light nuclei
\ $2\pi\Delta\left(  R_{P}+R_{T}\right)  \simeq300-600$ mb. Thus, the
probability $P$ violates unitarity (perturbation theory is invalid) if cross
sections of the order of 1 b are obtained.

Ref. \cite{FZ77} also introduced relativistic corrections to the nuclear
potential. This relativistic property is most easily seen within a folding
potential model for a nucleon-nucleus collision:
\begin{equation}
V\left(  \mathbf{r}\right)  =\int dr^{\prime3}\ \rho_{T}\left(  \mathbf{r}%
^{\prime}\right)  \ v_{NN}\left(  \mathbf{r-r}^{\prime}\right)  ,\label{fold}%
\end{equation}
where $\rho_{T}\left(  \mathbf{r}^{\prime}\right)  $ is the nuclear density of
the target. In the frame of reference of the projectile, the density of the
target looks contracted and particle number conservation leads to the
relativistic modification of eq. \ref{fold} so that $\rho_{T}\left(
\mathbf{r}^{\prime}\right)  \rightarrow\gamma\rho_{T}\left(  \mathbf{r}%
_{\perp}^{\prime}\mathbf{,\gamma}z^{\prime}\right)  $, where $\mathbf{r}%
_{\perp}^{\prime}$ is the transverse component of \textbf{r'} and
$\gamma=\left(  1-v^{2}/c^{2}\right)  ^{-1/2}$ is the Lorentz contraction
factor, with $v$ equal to the relative velocity of projectile and target. But
the number of nucleons as seen by the target (or projectile) per unit area
remains the same. In other words, a change of variables $z^{\prime\prime
}=\mathbf{\gamma}z^{\prime}$\ \ in the integral of eq. \ref{fold} seems to
restore the same eq. \ref{fold}. However, this change of variables also
modifies the nucleon-nucleon interaction $v_{NN}$. Thus, relativity introduces
non-trivial effects in a potential model description of nucleus-nucleus
scattering at high energies.

Colloquially speaking, nucleus-nucleus scattering at high energies is not
simply an incoherent sequence of nucleon-nucleon collisions. Since the
nucleons are confined within a box (inside the nucleus), Lorentz contraction
induces a collective effect: in the extreme limit $\gamma\rightarrow\infty$
all nucleons would interact at once with the projectile. This is often
neglected in pure geometrical (Glauber model) descriptions of nucleus-nucleus
collisions at high energies, as it is assumed that the nucleons inside
\textquotedblleft firetubes" scatter independently.

Assuming that the nucleon-nucleon interaction is of very short range so that
the approximation $\ v_{NN}\left(  \mathbf{r-r}^{\prime}\right)
=J_{0}\ \delta\left(  \mathbf{r-r}^{\prime}\right)  $\ can be used, one sees
from eq. \ref{fold} that $V\left(  \mathbf{r}\right)  $, the interaction that
a nucleon in the projectile has with the target nucleus, also has similar
transformation properties as the density: $V\left(  \mathbf{r}\right)
\rightarrow\mathbf{\gamma}V\left(  \mathbf{r}_{\perp}\mathbf{,\gamma}z\right)
$, i.e. $V\left(  \mathbf{r}\right)  $ transforms as the time-component of a
four-vector. In this situation, the Lorentz contraction has no effect
whatsoever in the diffraction dissociation amplitudes, described in the
previous sections within the eikonal approximation. This is because a change
of variables $Z^{\prime}=\mathbf{\gamma}Z$\ in the eikonal phases leads to the
same result as in the non-relativistic case, as can be easily checked from eq.
\ref{eikphase}. Of course, the delta-function approximation for the
nucleon-nucleon interaction means that nucleons will scatter at once, and
Lorentz contraction does not introduce any additional collective effect. This
is not the case for realistic interactions with finite range. Thus, nuclear
structure studied with high-energy nucleus-nucleus collisions is immensely
complicated by retardation effects and is not well understood.

\section{Emission of correlated pairs in peripheral reactions}

\begin{figure}[ptb]
\begin{center}
\includegraphics[
height=2.1698in,
width=2.9482in
]{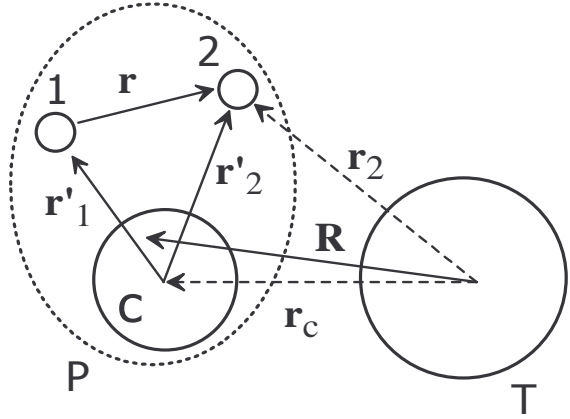}
\end{center}
\caption{Coordinates used in text for the three-body projectile interacting
with the target. the coordinate $\mathbf{r}_{1}$ is not shown for simplicity.
}%
\label{coll}%
\end{figure}

Lets us now consider the emission of correlated pairs in peripheral
collisions. The projectile is now a three-body system, with notation for the
coordinates as shown in figure \ref{coll}. Following the same arguments used
in section 2, the wavefunction of a three-body projectile in the initial and
final states is given by%
\begin{align}
\Psi_{i} &  =\phi_{i}\left(  \mathbf{r}_{1},\mathbf{r}_{2}\right)  \exp\left[
i\left(  \mathbf{K}_{c}\cdot\mathbf{r}_{c}+\mathbf{k}_{1}\cdot\mathbf{r}%
_{1}+\mathbf{k}_{2}\cdot\mathbf{r}_{2}\right)  \right]  \nonumber\\
\Psi_{f} &  =\phi_{f}\left(  \mathbf{r}_{1},\mathbf{r}_{2}\right)
S_{c}\left(  b_{c}\right)  S_{1}\left(  b_{1}\right)  S_{2}\left(
b_{2}\right)  \exp\left[  i\left(  \mathbf{K}_{c}^{\prime}\cdot\mathbf{r}%
_{c}+\mathbf{k}_{1}^{\prime}\cdot\mathbf{r}_{1}+\mathbf{k}_{2}^{\prime}%
\cdot\mathbf{r}_{2}\right)  \right]  \ ,
\end{align}
where now $\phi_{i,f}\left(  \mathbf{r}_{1},\mathbf{r}_{2}\right)  $ are the
initial and final intrinsic wavefunctions of the correlated nucleon-nucleon
pair as a function of \ their intrinsic coordinates $\mathbf{r}_{1},$
$\mathbf{r}_{2}$. Assuming that the nucleon mass is much smaller than that of
the core, one can replace $\mathbf{r}_{c}\simeq\mathbf{R}$, where \textbf{R}
is the center of mass of the projectile.

Following the same steps as before, a relation similar to eq.
\ref{dif2} can be obtained for the cross section for the energy
absorption by a correlated
pair (when final state interactions are neglected):%
\begin{align}
d\sigma &  =\frac{d^{3}q_{1}d^{3}q_{2}d^{2}Q}{\left(  2\pi\right)  ^{8}%
}\left\vert \int d^{3}r_{1}d^{3}r_{2}d^{2}b\ \phi_{f}^{\ast}\left(
\mathbf{r}_{1},\mathbf{r}_{2}\right)  \phi_{i}\left(  \mathbf{r}%
_{1},\mathbf{r}_{2}\right)  \right.  \ \nonumber\\
&  \times\left.  S\left(  b\right)  S_{1}\left(  b_{1}\right)  S_{2}\left(
b_{2}\right)  \exp\left[  i\left(  \mathbf{q}_{1}\cdot\mathbf{r}%
_{1}+\mathbf{q}_{2}\cdot\mathbf{r}_{2}+\mathbf{Q}\cdot\mathbf{b}\right)
\right]  \right\vert ^{2},
\end{align}
where $\mathbf{Q=K}_{c}^{\prime}-\mathbf{K}_{c}$. If the intrinsic nucleon
coordinates are denoted by $\mathbf{r}_{i}^{\prime}=\mathbf{r}_{i}-\mathbf{R}%
$, \ one has $b_{i}=\sqrt{b^{2}+r_{i}^{2}\sin^{2}\theta_{i}+2r_{i}b\sin
\theta_{i}\cos\left(  \phi-\phi_{i}\right)  }$.

The above relation can be used for the emission of the nucleon pair.
Neglecting final state interactions and assuming that the core is not observed
(i.e. integrating over $\mathbf{Q}$), one gets%
\begin{equation}
d\sigma=\frac{d^{3}q_{1}d^{3}q_{2}}{\left(  2\pi\right)  ^{6}}\int
d^{2}b\ \left\vert S\left(  b\right)  \right\vert ^{2}\ \left\vert \int
d^{3}r_{1}d^{3}r_{2}\ \phi_{i}\left(  \mathbf{r}_{1},\mathbf{r}_{2}\right)
S_{1}\left(  b_{1}\right)  S_{2}\left(  b_{2}\right)  \exp\left[  i\left(
\mathbf{q}_{1}\cdot\mathbf{r}_{1}+\mathbf{q}_{2}\cdot\mathbf{r}_{2}\right)
\right]  \right\vert ^{2}. \label{dsig5}%
\end{equation}

In order to proceed further one needs a model wavefunction for the correlated
pair, $\ \phi_{i}\left(  \mathbf{r}_{n},\mathbf{r}_{n^{\prime}}\right)  $. The
wavefunction used will have the form%
\begin{equation}
\phi_{i}\left(  \mathbf{r}_{1},\mathbf{r}_{2}\right)  =\phi_{\alpha}\left(
\mathbf{r}_{1}\right)  \phi_{\beta}\left(  \mathbf{r}_{2}\right)
f_{\mathrm{corr}}(\mathbf{r,r}_{c}\mathbf{)}\ ,\label{pairwf}%
\end{equation}
where $\mathbf{r}=\mathbf{r}_{1}-\mathbf{r}_{2}$,$\ \phi_{\alpha}\left(
\mathbf{r}\right)  =\phi_{nljm}\left(  \mathbf{r}\right)  $ are single
particle wavefunctions with quantum numbers $\alpha=nljm,$ and
$f_{\mathrm{corr}}\left(  \mathbf{r,r}_{c}\right)  $ is a function for the
nucleon pair distance \textbf{r}, which also depends on a two-particle
correlation parameter $\mathbf{r}_{c}$ so that $f_{\mathrm{corr}}\left(
\mathbf{r,r}_{c}\right)  \rightarrow0$ as $\mathbf{r}_{c}\rightarrow0$. The
effective correlation function $f_{\mathrm{corr}}\left(  \mathbf{r,r}%
_{c}\right)  $, the so-called Jastrow factor \cite{Ja55}, \ is a statistical
average of the Pauli correlation function \cite{BM69} and the correlation
function for the dynamical short-range \ (e.g., hard core) correlation.

As argued in ref. \cite{GWW58}, the true ground-state wave function of the
nucleus containing correlations coincide with the independent particle, or
Hartree-Fock wavefunction, for interparticle distances $r\geq r_{\mathrm{heal}%
}$, where $r_{\mathrm{heal}}\simeq1$ fm is the so-called \textquotedblleft
healing distance\textquotedblright. This behavior is a consequence of the
constraints imposed by the Pauli principle. Nucleons are kept apart at
short-distances, while for distances beyond several $K_{F}^{-1}$'s there is
little effect. Consequently, nucleon-nucleon collisions at short distances are
rare in nuclear matter, and because the strongest part of the interaction is
at short distances, the effective force \ between the nucleons is much less
than in free space. For example, if a nucleon in $^{16} $O felt the cumulative
sum of 16 nucleon-nucleon potentials, it would feel a potential of $\sim1400$
MeV; yet empirically it is known that the effective potential felt by the
nucleon in the middle of the nucleus is only $\sim40$-50 MeV deep.

Although, in general, the correlation function $f_{\mathrm{corr}}%
(\mathbf{r,r}_{c}\mathbf{)}$ may depend on the isospin and spin quantum
numbers of the two-body channel, we will assume for simplicity that it is a
plain, state independent, Jastrow factor \cite{Ja55}. The effects of
nucleon-nucleon correlations in nucleus-nucleus collisions have also been
studied in several works. For example, in ref. \cite{Ray79} short-range
correlations were shown to play an important role in nucleon-nucleus
collisions at intermediate and high energies.

\begin{figure}[ptb]
\begin{center}
\includegraphics[
height=2.3367in,
width=3.0831in
]{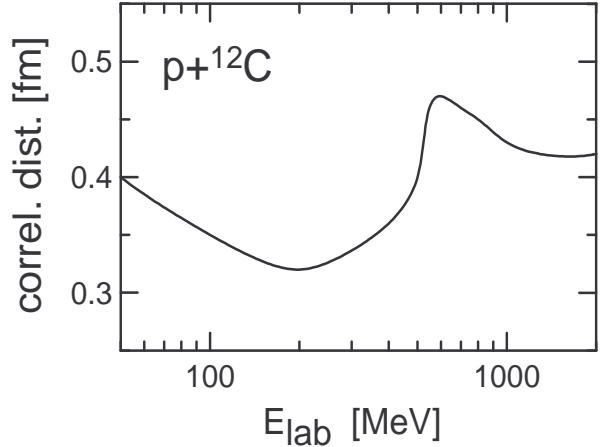}
\end{center}
\caption{Dependence of the short-range correlation distance, $r_{c}$, on the
proton energy in the reaction $p+^{12}C$.}%
\label{rcorr}%
\end{figure}

The two-particle correlation distance, $\mathbf{r}_{c}$, is a combination of
four contributions \cite{Ray79}%
\begin{equation}
r_{c}=r_{\mathrm{Pauli}}+r_{\mathrm{SRD}}+r_{\mathrm{PSR}}+r_{\mathrm{CM}},
\end{equation}
where $r_{\mathrm{Pauli}}$ is due to Pauli exclusion-principle correlations,
$r_{\mathrm{SRD}}$\ is related to short-range dynamical correlations,
$r_{\mathrm{PSR}}$\ is due to a combination of the Pauli and the short-range
dynamical term, and $r_{\mathrm{CM}}$\ is due to center-of-mass correlations
\cite{BF77}. An approximate set of expressions for each of these terms is
given by
\begin{align}
r_{\mathrm{Pauli}}  &  =\frac{1}{2}\left(  1-\frac{5}{A}+\frac{4}{A^{2}%
}\right)  \frac{3\pi}{10K_{F}}\frac{1}{1+\frac{8}{5}BK_{F}^{2}},\nonumber\\
r_{\mathrm{SRD}}  &  =\frac{1}{2}\left(  1-\frac{2}{A}+\frac{1}{A^{2}}\right)
\sqrt{\pi}\frac{b^{3}}{b^{2}+8B},\nonumber\\
r_{\mathrm{PSR}}  &  =\frac{1}{2}\left(  1-\frac{5}{A}+\frac{4}{A^{2}}\right)
\frac{3\pi}{10}\left(  K_{F}^{2}+\frac{5}{b^{2}}\right)  ^{-1/2}\left[
1+8B\left(  \frac{K_{F}^{2}}{5}+\frac{1}{b^{2}}\right)  \right]
^{-1}\nonumber\\
r_{\mathrm{CM}}  &  =\left(  1-\frac{2}{A}+\frac{1}{A^{2}}\right)  l_{c},
\end{align}
where $A$ and $K_{F}=\left(  1.5\pi^{2}\rho\right)  ^{1/3}\simeq1.36$
fm$^{-1}$ are the target number and the Fermi momentum of the target nucleus,
respectively. $b$ is a short-range dynamical correlation, $b\simeq0.4$ fm, $B$
is the finite-range parameter of the nucleon-nucleon elastic $t$-matrix,
$B\simeq0.62$ fm$^{2}$ (for collisions around 200 MeV/nucleon), and $l_{c}%
$\ is the effective \textquotedblleft correlation length", $l_{c}%
\simeq1.3\ A^{-5/6}$ fm$.$ For proton+$^{12}$C collisions at 200 MeV/nucleon
this set of parameters yields, $r_{\mathrm{Pauli}}\simeq0.3$ fm,
$r_{\mathrm{SRD}}\simeq0.01$ fm, $r_{\mathrm{PSR}}\simeq0.0016$ fm,
$r_{\mathrm{CM}}\simeq0.18$ fm, and $r_{c}\simeq0.5$ fm. This in fact
overestimates the correlation distance. A more detailed calculation, using the
parameters $B$, $b$ and $l_{c}$ from ref. \cite{Ray79} shows that $r_{c}$ has
an appreciable dependence on the collision energy, as shown in figure
\ref{rcorr} for protons incident on $^{12}$C. Thus, in nuclear reactions,
$r_{c}$ can vary substantially with the collision energy and with mass numbers.

The estimates done above show that the main contribution to the correlation
distance arises from the Pauli principle. Let us assume a correlation function
of the form%
\begin{equation}
f_{\mathrm{corr}}\left(  r\right)  =1-\exp\left[  -\frac{\left(
\mathbf{r}_{1}-\mathbf{r}_{2}\right)  ^{2}}{r_{c}^{2}}\right]  .\label{fcorr}%
\end{equation}
This correlation function implies that the pair wavefunction decreases for
small relative distances, $r=\left\vert \mathbf{r}_{1}-\mathbf{r}%
_{2}\right\vert \lesssim r_{c}$. The correlation function $f_{\mathrm{corr}%
}(\mathbf{r,r}_{c}\mathbf{)}$ goes to one for large values of $r$ and to zero
for $r\rightarrow0$. In nuclear structure calculations, the effect of
correlation, introduced by the function $f_{\mathrm{corr}}(\mathbf{r,r}%
_{c}\mathbf{)}$, becomes large when the correlation distance
parameter $r_{c}$ becomes large, and vice versa. Here, only the
effects of short-range correlations are studied and it would be
manifest in momentum distributions of highly energetic nucleons, as
discussed in the introduction, and explicitly shown in the next
section. It is important to notice that the Gaussian correlation
function, eq. \ref{fcorr}, is unrealistic. Indeed, the short-range
repulsion is at the origin of the decrease of the pair wavefunction
for small relative distances. At the same time, there will be an
increased probability to find the nucleon pair at medium
internucleon distances. A two-Gaussian parameterization is needed to
quantify this well-known effect of short-range correlations. For
simplicity, only the simple parameterization of eq. \ref{fcorr}\ is
used in this work.

Inserting eqs. \ref{pairwf} and \ref{fcorr} in eq. \ref{dsig5} and integrating
over the pair momenta, one gets%
\begin{align}
\sigma_{\mathrm{SR}} &  =\frac{\left(  C^{2}S\right)  _{lj}\left(
C^{2}S\right)  _{l^{\prime}j^{\prime}}}{\left(  2j+1\right)  \left(
2j^{\prime}+1\right)  }{\displaystyle\sum\limits_{m,\ m^{\prime}}}\int
d^{2}b\ \left\vert S\left(  b\right)  \right\vert ^{2}\ \nonumber\\
&  \times\int d^{3}r_{1}\ d^{3}r_{2}\ \left\vert \phi_{nljm}\left(
\mathbf{r}_{1}\right)  S_{1}\left(  b_{1}\right)  \phi_{n^{\prime}l^{\prime
}j^{\prime}m^{\prime}}\left(  \mathbf{r}_{2}\right)  S_{2}\left(
b_{2}\right)  f_{\mathrm{corr}}(\mathbf{r,r}_{c}\mathbf{)}\right\vert
^{2}\ .\label{cor}%
\end{align}
The cross section has been averaged over the initial magnetic quantum numbers
of the nucleons. If the correlation function were equal to the unity, the
integrand would be the product of the probabilities to remove an uncorrelated
nucleon, with quantum numbers $nljm$. The later probability is given by $\int
d^{3}r\ \left\vert \phi_{nljm}\left(  \mathbf{r}\right)  \ S_{i}\left(
b\right)  \right\vert ^{2}$.

The spectroscopic factors in eq. \ref{cor} have a complex dependence on the
angular momenta of the nucleon pair. The correlations arising from angular
momentum coupling have been studied in ref. \cite{Jef04}. Here we will assume
a simple combinatorics so that $\left(  C^{2}S\right)  _{lj}=n\left(
n-1\right)  /2$, where $n$ is the number of nucleons in the valence shell.

We see from the equations above that the cross section for the emission of a
correlated pair is smaller than that for the emission of independent
particles, since $f_{\mathrm{corr}}(\mathbf{r,r}_{c}\mathbf{)}\leq1$. Most
part of the integrand will have $f_{\mathrm{corr}}(\mathbf{r,r}_{c}%
\mathbf{)}\sim1$, except for the small region of volume $\mathcal{N}r_{c}^{3}%
$, where $\mathcal{N}$ is a number of order of one. Conservative estimates
(using $r_{c}=0.3-1$ fm), imply that the cross section for emission of a
correlated pair could not exceed 100 mb, in contrast to the results obtained
in refs. \cite{FZ77,Fes80}. We will show this for specific reactions in the
following section.

\section{Results and discussions}

The numerical calculations have been carried out for the systems $^{12}%
$C+$^{12}$C at 250 MeV/nucleon and $^{11}$Li+$^{9}$Be at 287 MeV/nucleon. In
both cases, there are some experimental data available for two-nucleon
removal. This also allows for the study of the influence of a halo
wavefunction ($^{11}$Li) in the results. The wavefunctions were calculated by
using a Woods-Saxon potential with a spin-orbit and Coulomb potential,
\begin{equation}
V\left(  r\right)  =U_{r}\left(  r\right)  +U_{s}\left(  r\right)
+U_{C}\left(  r\right)  ,
\end{equation}
where%
\begin{equation}
U_{r}\left(  r\right)  =V_{r}\left(  1+e^{\rho_{r}}\right)  ^{-1}%
,\ \ \ \ \ \ \ \ \ U_{s}\left(  r\right)  =V_{s}\left(  \mathbf{l\cdot
s}\right)  \frac{\left(  2\ \mathrm{fm}^{2}\right)  }{r}\frac{d}{dr}\left(
1+e^{\rho_{s}}\right)  ^{-1},
\end{equation}
$U_{C}\left(  r\right)  $ is the potential for a uniformly charged sphere with
charge $Z-1$ ($Z$, for neutrons) and radius $R_{C}$, and $\rho_{i}=\left(
r-R_{i}\right)  /a_{i}$.

For protons in the $1p_{3/2}$ orbital of $^{12}$C the separation energy is
15.96 MeV (the two-proton separation energy is 27.18 MeV), which can be
reproduced with the parameters $V_{r}=-57.41$ MeV, $V_{s}=-6.0$ MeV,
$R_{r}=R_{C}=R_{s}=3.011$ fm, $a_{r}=0.52$ fm and $a_{s}=0.65$ fm.

The reactions and structure of the two-neutron halo nucleus $^{11}$Li have
attracted much interest. It is a Borromean system in the sense that although
the three-body system, consisting of $^{9}$Li and two neutrons, forms a bound
state, none of the possible two-body subsystems have bound states. Hence the
stability of $^{11}$Li is brought about by the interplay of the core-neutron
and the neutron-neutron interactions, which must lead to a strongly correlated
wave function with the two neutrons spatially close together. For the
calculation here we will approximate the $^{11}$Li ground state by an inert
$^{9}$Li core coupled to a neutron pair in a $\left(  2s1/2\right)  ^{2}$
state, although the most probable configuration is an admixture of neutron
pairs in $\left(  2s1/2\right)  ^{2}$, $\left(  1p1/2\right)  ^{2}$, and
$\left(  1d5/2\right)  ^{2}$ states \cite{Si99,BH04}. However, the former
assumption allows for a simpler calculation of the correlated-pair emission.
The potential parameters are adjusted to obtain the single-particle wave
functions, reproducing the effective neutron separation energies. The
two-neutron separation energy is 0.3 MeV. From the systematics in Fig. 6 of
\cite{Han01}, the estimated $^{10}$Li average excitation energies is 0.2 MeV
for the single-particle state. Taking this value for two-neutron coupling to
the $^{9}$Li core, one arrives at an effective neutron-separation energy of
0.5 MeV. This binding energy for the n+$^{10}$Li system can be reproduced with
the potential parameters $V_{r}=-42.93$ MeV,$V_{s}=-6.0$ MeV, $R_{r}%
=R_{C}=R_{s}=3.25$ fm and $a_{r}=a_{s}=0.65$ fm.

The single-particle wavefunctions obtained in this way were used in eq.
\ref{dsig5}\ to calculate the momentum distributions of the correlated pair.
The integrals in eq. \ref{cor} were performed using a method similar to that
described in the appendix of ref. \cite{BH04}. The S-matrices (and optical
potentials) were calculated by using the \textquotedblleft t-$\rho\rho$"
interaction, as described in refs. \cite{Ray79,RHB91}. This is the same
approximation used in ref. \cite{HT04}.

In heavy ion physics it is common to define a correlation function by means of%
\begin{equation}
C\left(  \mathbf{q}_{1},\mathbf{q}_{2}\right)  =\left(  \frac{d\sigma}%
{d^{3}q_{1}d^{3}q_{2}}\right)  \left/  \left(  \frac{1}{\sigma}\frac{d\sigma
}{d^{3}q_{1}}\frac{d\sigma}{d^{3}q_{2}}\right)  \right.  ,
\end{equation}
where the cross sections in the denominator are for the emission of a single
nucleon. An accurate measurement of $r_{c}$ requires the measurement of this
correlation function for back-to-back (or nearly) pair emission. Until now,
heavy ion data refer mainly to small relative momentum transfers. Data would
only be interesting for the present purposes if the triggering conditions were
changed and if special attention was paid to back-to-back emission. In the
present work, $d\sigma/dq_{1}dq_{2}$, instead of $C\left(  \mathbf{q}%
_{1},\mathbf{q}_{2}\right)  $, will be used for the study of emission of
correlated nucleons.

\begin{figure}[ptb]
\begin{center}
\includegraphics[
height=3.5552in,
width=4.4425in
]{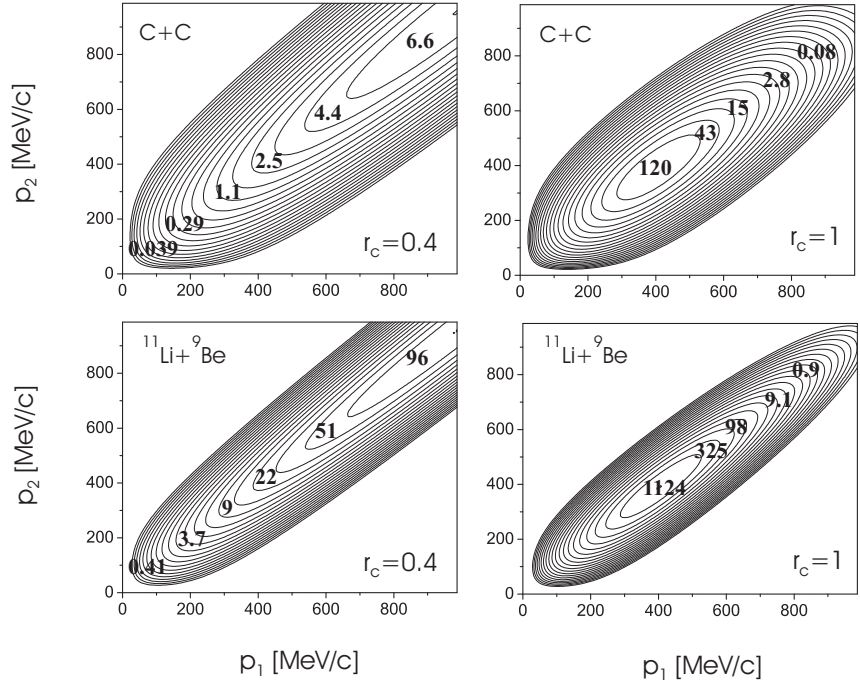}
\end{center}
\caption{Contour plots of $d\sigma/dq_{1}dq_{2}$ for the collision $^{12}%
$C+$^{12}$C at 250 MeV/nucleon (upper panels) and $^{11}$Li+$^{9}$Be at 287
MeV/nucleon (lower panels), as a function of $p_{1}=\hbar q_{1}$ and
$p_{2}=\hbar q_{2}$..The left (right) panels are for $r_{c}=0.4\ (1)$ fm. The
numbers in the plots indicate the cross section, $d\sigma/dq_{1}dq_{2}$,\ in
units of 10$^{-5}$ [mb/(MeV/c)]$^{2}$.}%
\label{contour}%
\end{figure}

In figure \ref{contour}\ contour plots for $d\sigma/dq_{1}dq_{2}$ are
presented for the collision $^{12}$C+$^{12}$C at 250 MeV/nucleon and $^{11}%
$Li+$^{9}$Be at 287 MeV/nucleon, as a function of $p_{1}=\hbar q_{1}$ and
$p_{2}=\hbar q_{2}$. The nucleons are assumed to be emitted back-to-back, the
nucleon 1 at 0$^{o}$ and nucleon 2 at 180$^{o}$ with respect to the beam axis,
respectively. The upper panels are for the C+C collision, while the lower
panels are for the Li+Be collisions. The left (right) panels are for
$r_{c}=0.4$ fm ($r_{c}=1$ fm). The numbers in the plot indicate the cross
section, $d\sigma/dq_{1}dq_{2}$,\ in units of 10$^{-5}$ [mb/(MeV/c)]$^{2}$.
One notices a strong correlation between the nucleon momenta, resulting from
the phonon relationship, eq. \ref{phonon} . The effect of the phonon
dispersion relation is to produce a ridge in the cross section.

\bigskip

To obtain a greater physical insight, I will now use the PWBA approximation as
in eq. \ref{TDWBA0} (with $S(b)=1$), so that, instead of the integrals in eq.
\ref{dsig5}, one needs now to calculate
\begin{equation}
T_{\mathrm{(PWBA)}}=\int d^{3}r_{1}\ d^{3}r_{2}\ \phi_{i}\left(
\mathbf{r}_{1},\mathbf{r}_{2}\right)  \left[  V_{1}\left(  \mathbf{r}%
_{1}\right)  +V_{2}\left(  \mathbf{r}_{2}\right)  \right]  \exp\left[
i\left(  \mathbf{q}_{1}\cdot\mathbf{r}_{1}+\mathbf{q}_{2}\cdot\mathbf{r}%
_{2}\right)  \right]  .\label{DWBApair}%
\end{equation}
Let us assume that the potentials $V_{1,2}$ are given by Gaussian functions,
i.e. $V_{1,2}=V_{1,2}^{(0)}\ \exp\left(  -r_{1,2}^{2}/\lambda^{2}\right)  $
and similarly for the wavefunctions, i.e. $\phi_{\alpha,\beta}=N\exp\left(
-r_{1,2}^{2}/\Delta^{2}\right)  $. In this case it is straightforward to
perform analytically all the integrals in eq. \ref{DWBApair}. If one further
assumes that the correlation distance, $r_{c}$, is much smaller than the
dimensions of the uncorrelated wavefunctions and of the potential, i.e. if
$r_{c}\ll\Delta,\lambda$, one gets
\begin{align*}
T_{\mathrm{(PWBA)}} &  =\left(  V_{1}^{(0)}+V_{2}^{(0)}\right)  \frac{\left(
\pi N^{2}r_{c}\Delta/\sqrt{\mathcal{A}}\right)  ^{3}}{8}\left[  \exp\left(
-\frac{q_{1}^{2}r_{c}^{2}}{4}\right)  +\exp\left(  -\frac{q_{2}^{2}r_{c}^{2}%
}{4}\right)  \right]  \\
&  \times\exp\left[  -\frac{\Delta^{2}}{16\mathcal{A}}\left\vert
\mathbf{q}_{1\perp}+\mathbf{q}_{2\perp}\right\vert ^{2}\right]  \exp\left[
-\frac{\Delta^{2}}{16\mathcal{A}}\left(  q_{1z}+q_{2z}-\frac{\omega}%
{v}\right)  ^{2}\right]  ,
\end{align*}
where $B$ is the separation energy of the pair and $\mathcal{A=}\left(
1+\Delta/\lambda\right)  /2$. Therefore, the cross section is given by\
\begin{align}
\frac{d\sigma}{dq_{1}dq_{2}} &  \propto q_{1}^{2}q_{2}^{2}\exp\left[
-\frac{\Delta^{2}}{8\mathcal{A}}\left(  q_{1z}+q_{2z}-\frac{\omega}{v}\right)
^{2}\right]  \ \nonumber\\
&  \times\exp\left[  -\frac{\Delta^{2}}{8\mathcal{A}}\left\vert \mathbf{q}%
_{1\perp}+\mathbf{q}_{2\perp}\right\vert ^{2}\right]  \ \left[  \exp\left(
-\frac{q_{1}^{2}r_{c}^{2}}{4}\right)  +\exp\left(  -\frac{q_{2}^{2}r_{c}^{2}%
}{4}\right)  \right]  ^{2}.\label{momdis}%
\end{align}

Now one can easily understand the physics in figure \ref{contour}\ by
identifying the terms of the above equation. The first term is due to
conservation of the momentum along the beam direction, which yields the
dispersion relation, eq. \ref{phonon}. Note that in the derivation of eq.
\ref{dsig5} it was assumed that the core recoils with the same momentum, i.e.
$Q_{Z}\simeq\omega/v=-\left(  q_{1z}+q_{2z}\right)  $. The second term is due
to elastic scattering of the pair on the target in the direction transverse to
the beam. In eq. \ref{momdis} both the first and the second terms imply that
the momentum distribution of correlated pair is such that $q_{1}+q_{2}%
=\omega/v$, i.e. $q_{1}\simeq-q_{2},$ as expected for small $r_{c}$. Also
according to these terms, the distribution is smeared by the range of the
independent wavefunctions of the pair, i.e. $\left\langle q_{1,2}%
^{2}\right\rangle \simeq1/\Delta^{2}$. However, the last term implies a
smearing, or spreading, of the momentum distribution by a much larger factor
(assuming $r_{c}\ll\Delta$), i.e. $\left\langle q_{1,2}^{2}\right\rangle
\simeq1/r_{c}^{2}$. This explains all physics presented in figure
\ref{contour}. The second exponential term in eq. \ref{momdis}\ plays no role
in the results presented in figure \ref{contour}, since it is identical to one.

As discussed above, the location of the ridges in figure \ref{contour} is a
kinematical property of the phonon absorption mechanism, which is independent
of the collision energy. Thus, it should be observable in intermediate energy
collisions ($E_{lab}\simeq100$ MeV/nucleon), as well as in relativistic
collisions. One also observes that the momentum distributions are narrower for
correlated-pair emission from a halo nucleus. This is due to the low binding
energy, which yields an extended wavefunction of the nucleons in the halo.
This is also seen from the first exponential term of eq. \ref{momdis}, since
the two-proton separation energy for $^{12}$C is 27.16 MeV, while the
two-neutron separation energy is 0.3 MeV. The effective value of $\Delta$\ is
much smaller for the first case, leading to a larger spreading of the momentum
distributions. However, the last term in eq. \ref{momdis} is still the
dominant one leading to a small overall effect on the momentum distribution,
as shown next.

\begin{figure}[ptb]
\begin{center}
\includegraphics[
height=2.1119in,
width=3.2465in
]{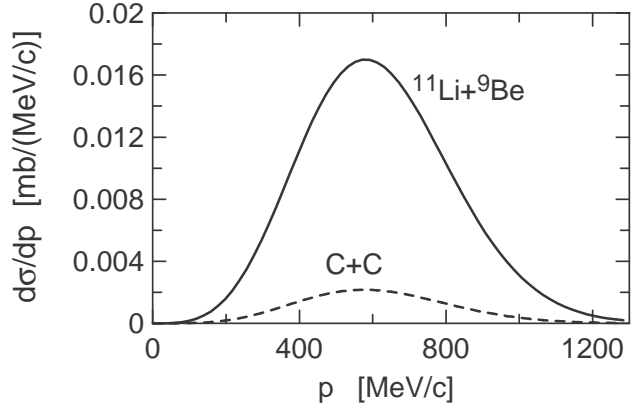}
\end{center}
\caption{Singles spectra for the momentum distribution of a nucleon due to the
correlated-pair mechanism.}%
\label{singles}%
\end{figure}

It would be interesting to try to observe the contribution of the emission of
correlated pairs in singles spectra. This can be obtained by integrating
$d\sigma/dq_{1}dq_{2}$ over one of the two nucleon momenta. This is shown in
figure \ref{singles}\ for C+C and Li+Be collisions, using $r_{c}=0.7$ fm. One
observes that the peak in the singles spectra occurs at $p\simeq\hbar/r_{c},$
as expected from the arguments presented above. This should be visible in the
spectra of nucleons from knockout reactions as a bump at high nucleon momenta.
The position of the bump would be a direct reading of the short-range
correlation distance. Notice, however, that such a bump could not be
noticeable because it is superposed to a large background of knockout nucleons
from stripping reactions. Only by doing a measurement of back-to-back pair
emission, this signature of short-range pair-correlations could be assessed.

The total cross section for the emission of correlated pairs arising
from short-range correlations can be calculated from eq. \ref{cor}.
Assuming $r_{c}=0.7$ fm, the total cross section for the emission of
high-energy correlated pairs in C+C collisions at \ 250 MeV/nucleon
is $\sigma _{\mathrm{corr}}=0.61$ mb. The experimental value for
two-proton knockout in this collision is $5.88\pm9.70$ mb. For
$^{11}$Li+$^{9}$Be at 287 MeV/nucleon the correlated pair cross
section is $\sigma_{\mathrm{corr}}=4.1$ mb. These
cross sections are much smaller than those obtained in refs. \cite{FZ77,Fes80}%
. For reasons which were explained in the paragraph preceding eq. \ref{fold},
the results obtained here are much more reasonable. These cross sections are
also much smaller than those for one-nucleon knockout reactions (see, e.g.
ref. \cite{HT04}). They are also only one of the contributions (i.e. only from
diffraction dissociation) of the two-proton removal cross section. Another
contribution (stripping) has not been considered here. Stripping would not
contribute to back-to-back nucleon emission, with nearly zero total momentum
of the pair, but is responsible for the largest part of the two-nucleon
knockout total cross section.

In conclusion, I have shown that when a projectile reacts with a light nuclear
target, the short-range correlations contribute to the emission of high-energy
nucleons which can be visible in measurements of back-to-back emission of
nucleon pairs. More experiments and also the development of a more complete
reaction theory are interesting challenges. The theoretical results suggest
that the pattern and absolute magnitudes of the partial cross sections can
provide specific information on the detailed nature of the states involved.
This is particularly important in the case of reactions involving neutron-rich
and proton-rich nuclei, far from the stability valley, for which only nuclear
reactions are presently capable of probing their internal structure. The
results presented here will be valuable as a guide to extend these studies
towards drip line nuclei and look for effects which cannot be probed in
($e,e^{\prime}$) scattering due to the lack of experimental facilities of
electron scattering on drip line nuclei.

\bigskip\bigskip

I would like to thank Angela Bonaccorso, Kai Hencken and Ian Thompson for
beneficial discussions. This work was supported by the U.\thinspace
S.\ Department of Energy under grant No. DE-FG02-04ER41338.

\end{document}